\documentclass[5p,times]{elsarticle}

\journal{Systems and Control Letters}

\usepackage{amsmath} 
\usepackage{amssymb}  
\usepackage{amsthm}
\usepackage{physics}
\usepackage{graphicx}

\usepackage[margin=2mm]{subfig}
\usepackage{color}
\usepackage{tikz}
\usepackage{pgfplots}
\usepackage{epstopdf}
\usepackage{float}   
\usepackage{hyperref}

\theoremstyle{remark}

\newcommand{\SO}{\ensuremath{\mathrm{SO(3)}}}

\newcommand{\T}{^{\mbox{\small T}}}
\newcommand{\so}{\ensuremath{\mathfrak{so}(3)}}

\newcommand{\bi}{\begin{itemize}}
	\newcommand{\ei}{\end{itemize}}

\newcommand{\bR}{\ensuremath{\mathbb{R}}}

\newcommand{\bbm}{\begin{bmatrix}}
	\newcommand{\ebm}{\end{bmatrix}}
\newcommand{\matl}{\left[ \begin{array}}
	\newcommand{\matr}{\end{array} \right]}
\newcommand{\be}{\begin{equation}}
\newcommand{\ee}{\end{equation}}
\newcommand{\bea}{\begin{eqnarray}}
\newcommand{\eea}{\end{eqnarray}}
\newcommand{\beas}{\begin{eqnarray*}}
	\newcommand{\eeas}{\end{eqnarray*}}
\newcommand{\nn}{\nonumber}

\newcommand{\di}{\mathrm{d}}

\newcommand{\lan}{\langle}
\newcommand{\ran}{\rangle}

\renewcommand{\qedsymbol}{$\blacksquare$}

\begin{document}	
\begin{frontmatter}

 \title{\LARGE \bf Geometric PID-type attitude tracking control on $\SO$}

 \author{Hossein Eslamiat}
 \ead{heslamia@syr.edu}
 
 \author{Ningshan Wang} 
 \ead{nwang16@syr.edu}
 
 \author{Amit K. Sanyal}
 \ead{aksanyal@syr.edu}

\address{Department of Mechanical and Aerospace Engineering, Syracuse University, Syracuse, NY 13244, USA}

\begin{abstract}     
This article develops and proposes a geometric nonlinear proportional-integral-derivative (PID) type tracking 
control scheme on the Lie group of rigid body rotations, $\SO$. Like PD-type attitude 
tracking control schemes that have been proposed in the past, this PID-type control scheme exhibits almost 
global asymptotic stability in tracking a desired attitude profile. The stability of this PID-type tracking control 
scheme is shown using a Lyapunov analysis. A numerical simulation study 
demonstrates the stability of this tracking control scheme, as well as its robustness to a disturbance torque. In addition, a numerical comparison study shows the effectiveness of the proposed integrator term. 
\end{abstract}
\end{frontmatter}
\section{Introduction}\label{sec:intro}
Classical PID control schemes are widely used in practice and have several applications due to their ease of design and tunable properties. PID controllers create a control input based on a tracking error, which is the difference between the actual output and a desired (reference) output. This control input has three terms: one proportional to the error, one proportional to the time integral of the error, and another term proportional to the time derivative of the error. Using PID feedback has the advantage of eliminating steady state error (using an integral term) and reducing oscillations (using a derivative term)~\cite{Franklin}. In addition, when the mathematical model of a plant is not known and hence analytical design methods cannot be used, PID controllers prove to be very useful~\cite{ogata}. The popularity of PID controllers can be attributed partly to their good performance in a wide range of operating conditions and partly to their functional simplicity~\cite{Dorf}.  
Some PID controllers have been proposed for rigid body attitude tracking by utilizing local coordinates or quaternions, such as the ones in~\cite{wen,CLi2012,JSu,Subbaro}, however, these suffer from singularities or unwinding~\cite{rotmat2011}. Unwinding occurs when in response to certain initial conditions, a closed loop trajectory undergoes a homoclinic-like orbit that initiates near the desired attitude equilibrium. For more details on unwinding, see~\cite{BhatBernsteinUnwinding,Mayhew}.

Geometric mechanics is the interface for applying geometric control to mechanical systems. This approach results in conserving features of the configuration space without the need of local coordinates or parameterization.  An early work extending classical PD control to mechanical systems evolving on configuration manifolds is~\cite{koditschek}, where PD-type control was used to stabilize a desired configuration. It is worth noting that controls for such systems are defined on the tangent space of the configuration manifold. In subsequent years, others have proposed various geometric PD-type controllers, such as in~\cite{Bullo,TLee2010,DHS2006,Reza}. If there is a bounded parameter error or disturbance, a geometric PD controller can guarantee global boundedness of tracking errors, although, they might not converge to zero. By choosing sufficiently large PD gains, the errors can be made arbitrarily small. However this can result in amplifying undesirable noise, saturating actuators, and requiring large control effort. This, alongside the added robustness, motivates adding an integrator to a PD type controller.

Research on geometric PID-type control includes~\cite{Sarlette}, in which the authors consider control of a mechanical system on a Lie group.
They propose an integral action, evolving on the Lie group, to compensate the drift resulting from a constant bias in velocity and torque inputs. However, they assume a constant time-invariant bias, and only discuss feedback stabilization and not the feedback tracking problem.
The work in~\cite{FGoudarziECC} defines an integral term by putting the derivative of integral error equal to the intrinsic gradient of the error function \emph{plus} a velocity error term. However, since the derivative of the integral term is not on the tangent space, the integrator is nonintrinsic. Therefore, the integrator depends on the coordinates chosen for the Lie algebra of the Lie group, unlike the intrinsic PID controller proposed in~\cite{DHSandBerg}.
A more recent work~\cite{Banavar} considers the tracking problem and proposes a PID controller for a rigid body with internal rotors. This builds on the previous PID controller designed in~\cite{DHSandBerg}, where it is shown an intrinsic (geometric) integral action ensures that tracking errors converge to zero, in response to constant velocity commands. Following up on this prior work,~\cite{Banavar} develops an intrinsic PID controller on $\SO$ for attitude tracking applications. The proposed PID-type controller and tracking algorithm can work in conjunction with trajectory generation algorithms such as in~\cite{MuellerD,Had,cuthow12,EsLi,mellinger2011minimum,Ingersoll16,LiEslamiat}. A generated trajectory can be considered as the desired trajectory and be tracked by the algorithm presented in this paper.

This paper is organized as following. Section \ref{Sec:PrForm} formulates the problem by introducing coordinate frames used, reference attitude trajectory and attitude dynamics. Section \ref{Sec:TrackErr} discusses tracking error kinematics and dynamics. Section \ref{sec: MainRes} proposes the geometric PID controller, along with its stability proof. Section \ref{sec:NumRes} is dedicated to a Lie group variational integrator (LGVI) discretization and numerical validation, and comparison with a PD type controller and discussion of disturbance-free case. Finally, section \ref{sec:conc} concludes the paper and gives directions for future work.

\section{Problem Formulation}\label{Sec:PrForm}
The treatment in this paper is general and can be applied to vehicles modeled as rigid bodies, e.g., spacecraft, unmanned underwater vehicles and unmanned aerial vehicles like quadrotors. 
\subsection{Coordinate frames}
The two coordinate systems used to define the attitude of a rigid body are inertial and body-fixed coordinates. Attitude of the vehicle is defined as the rotation from body-fixed frame to inertial frame and is denoted $R\in\SO$. For attitude tracking, we also define a desired attitude trajectory in time, denoted $R_d (t)$. In addition, we denote by $\Omega$ the angular velocity and $\Omega^d$ denotes the desired angular velocity.
\subsection{Reference Attitude Generation}
The desired attitude trajectory for the rigid body is assumed to be generated and available a priori. 
As an example for rotorcraft unmanned aerial vehicles (UAVs) the desired attitude trajectory can be generated 
from the position trajectory, by using the known dynamics model and actuation. Let $m$ and $J$ denote mass and inertia of a rigid body, respectively. The rotational dynamics 
of the rigid body is given by:
\begin{align}\label{Dyn}
\dot{R}&=R\,\Omega^\times, \\
J\dot{\Omega}&=J\Omega \times \Omega+\tau,\label{Dyn2}
\end{align}
where $\tau$ is the input torque and the cross map:$ (\cdot)^{\times}: \mathbb{R}^3\rightarrow \SO$ is given by~\cite{SanyalChyba}:
\begin{align*}
    x^\times= \bbm x_1\\ x_2\\ x_3\ebm^\times= \bbm 0 & -x_3 & x_2\\ x_3 & 0 & -x_1\\ 
-x_2 & x_1 & 0\ebm.
\end{align*}
\section{Tracking error kinematics and dynamics on TSO(3)}\label{Sec:TrackErr}
The attitude tracking error is defined by~\cite{SanyalChyba}:
\begin{align}\label{Q}
Q=R_d\T R.
\end{align}
Taking the time derivative results in:
\begin{align}
\begin{split}
\dot{Q}&=\dot{R}_d\T R+R_d\T\dot{R}=(R_d{(\Omega^d)}^{\cross})\T R+R_d\T R\Omega^{\cross} \\
&=R_d\T R\Omega^{\cross}-{(\Omega^d)}^{\cross}R_d\T R=Q\Omega^{\cross}-{(\Omega^d)}^{\cross}Q \\
&=Q(\Omega-Q\T\Omega^d)^{\cross}=Q\omega^{\cross},
\end{split} \label{Qdot}
\end{align}
where $\omega=\Omega-Q\T\Omega^d$ is the angular velocity tracking error. As a result, 
\begin{align}\label{JOmegadot}
    J\dot{\Omega}&=J\frac{d}{dt}(\omega+Q\T\Omega^{d})=J(\dot{\omega}+{\dot{Q}}\T\Omega^d+Q\T\dot{\Omega}^d)\nn\\
    &=J(\dot{\omega}+{(Q\omega^{\cross})}\T\Omega^d+Q\T\dot{\Omega}^d)\nn\\
    &=J(\dot{\omega}+Q\T\dot{\Omega}^d-\omega^{\cross}Q\T\Omega^d)
\end{align}
\begin{figure}[H]
\centering
\includegraphics[width=.95\columnwidth]{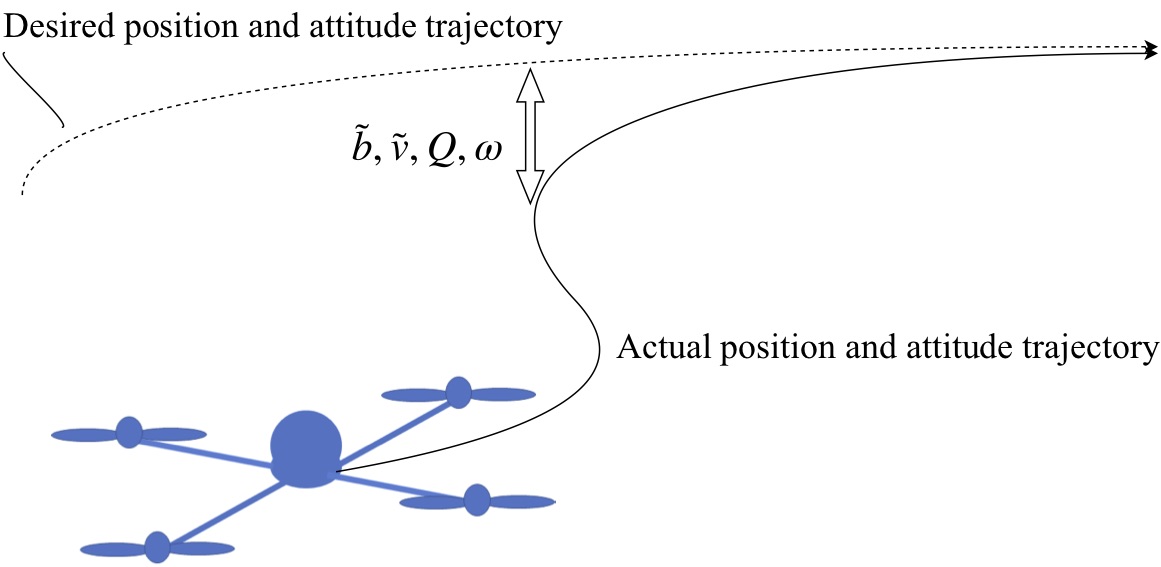}
\caption{Tracking errors}
\label{Fig:tracking errors}
\end{figure}
Figure (\ref{Fig:tracking errors}) shows a schematic of tracking errors for a quadrotor UAV model, as the 
difference between the desired trajectory and actual trajectory. In this figure, $\Tilde{b}$ denotes position tracking error and $\Tilde{v}$ denotes translational velocity tracking error, both defined in inertial frame.

\section{Main Result} \label{sec: MainRes}
\noindent \emph{Lemma 1}\\
\emph{Let} $\left\langle X,Y\right\rangle$ \emph{denote} $\tr (X\T Y)$ \emph{and $e_1$, $e_2$, $e_3$  be unit vectors in $x$, $y$, $z$ directions, respectively. Let $I$ denote the $3\times3$ identity matrix and $K$ be:}
\begin{align*}
    K=\bbm 
    k_1 & 0 & 0\\ 0 & k_2 & 0\\ 
0 & 0 & k_3\ebm \text{  \emph{where $k_i$ are distinct positive scalars,} }
\end{align*}
\emph{and define $S_K(Q)$ as:}
\begin{align}\label{sKQ}
S_K(Q)=\sum\limits_{i=1}^{3}k_i(Q\T e_i)\times e_i, \phantom{x} 
\end{align}
\emph{such that} $\frac{\di}{\di t}\lan K,I-Q\ran=\omega\T S_K(Q)$. \emph{Then} $\lan K,I-Q\ran$ \emph{is a Morse function on \SO.}
\\ The proof of Lemma 1 is given in~\cite{IZADIAtt} and is omitted here for brevity.
 \\  \\
\noindent \emph{Theorem 1} \\
\emph{Let $k_P, k_D, k_I\in\bR^+$ denote proportional, derivative and integrator feedback gains, respectively, and let $S_K(Q)$ be defined as in Lemma 1. Let $F_I\in\bR^3\simeq\so$ be the proposed integrator term given by:
\begin{equation}\label{jayom}
\begin{aligned}
J\dot{F}_I=-k_p S_K(Q)-k_D\omega,\;\ F_I(0)=0.
\end{aligned}
\end{equation}
Considering attitude dynamics of a rigid body as in equations (\ref{Dyn})-(\ref{Dyn2}), then the following control law: 
\begin{align}\label{eq:tauthm}
    \tau=&k_IF_I-k_PS_K(Q)-k_D\omega\nn\\
    &+J(Q\T {\dot{\Omega}}^d-\omega^{\cross} Q\T \Omega^d)-J\Omega\cross Q\T{\Omega}^d,
\end{align}
leads to asymptotically stable tracking of $(R_d,\Omega^d)$, where $(Q,\omega)$ are tracking errors given by (\ref{Q})-(\ref{Qdot}).} 
\\ \\
\emph{Proof} \\
Let $V:\SO\times\bR^3\to\bR^+$ be a Lyapunov candidate given by:
\begin{align}\label{eq:LypFcn}
   V=\bar{V}+k_p\left\langle K,I-Q\right\rangle,
\end{align}
where
\begin{align}\label{Vbar}
  \bar{V}=\frac{1}{2}\left\{(F_I-\omega)\T J(F_I-\omega)+\omega\T J\omega\right\}.
\end{align}
Note that by Lemma 1, $\left\langle K,I-Q\right\rangle$ is a Morse function on $\SO$, so is
$V$.
By comparing equation (\ref{JOmegadot}) and dynamics equation (\ref{Dyn}) we get:
\begin{align}
    J\dot{\omega}+J(Q\T\dot{\Omega}^d-\omega^{\cross}Q\T\Omega^d)=J\Omega \times \Omega+\tau,
\end{align}
then,
\begin{align}\label{eq:Jodot}
    J\dot{\omega}=J\Omega \times \Omega+J(\omega^{\cross}Q\T\Omega^d-Q\T\dot{\Omega}^d)+\tau.
\end{align}
As a result,
\begin{align}\label{oJodot}
   \omega\T J\dot{\omega}=\omega\T \lbrack J\Omega \times \Omega+J(\omega^{\cross}Q\T\Omega^d-Q\T\dot{\Omega}^d)+\tau\rbrack.
\end{align}
Consider equation (\ref{eq:LypFcn}), according to Lemma 1, by taking time derivative of $V$ we get:
\begin{align}\label{Vdot}
   \dot{V}=\dot{\bar{V}}+k_p\omega\T S_K(Q).
\end{align}
Taking time derivative of $\bar{V}$ in equation (\ref{Vbar}) gives 
\begin{align}\label{Vbardot}
  \dot{\bar{V}}=(F_I-\omega)\T J(\dot{F_I}-\dot{\omega})+\omega\T J\dot{\omega},
\end{align}
where $J\dot{\omega}$ is as expressed in equation (\ref{eq:Jodot}), and $F_I$ as in equation (\ref{jayom}).
Consider equation (\ref{oJodot}), and set:
\begin{align}
    &J\Omega\cross\Omega+J(\omega^{\cross} Q\T \Omega^d - Q\T {\dot{\Omega}}^d)+{\tau}=\nn\\
    &k_I F_I+J\Omega\cross\omega-k_P S_K(Q)-k_D\omega.
\end{align}
This gives the control torque as:
\begin{align}\label{eq:tau}
    \tau=&k_IF_I-k_PS_K(Q)-k_D\omega\nn\\
    &+J(Q\T {\dot{\Omega}}^d-\omega^{\cross} Q\T \Omega^d)-J\Omega\cross Q\T{\Omega}^d.
\end{align}
Then as a result
\begin{align}\label{eq:oTJodot}
    \omega\T J\dot{\omega}=-k_P\omega\T S_K(Q)-k_D\omega\T\omega+k_I\omega\T F_I.
\end{align}
Using equation (\ref{eq:oTJodot}), and replacing equation (\ref{Vbardot}) in (\ref{Vdot}) results in:
\begin{align}
    \dot{V}=-k_I F_I\T F_I+2k_I\omega\T F_I
    -k_D\omega\T\omega.
\end{align}
By setting $k_D=k_I+k_{DI}$ where $k_{DI}>0$, we get:
\begin{align}\label{eq:VdotLess0}
   \dot{V}=
   -k_{DI}\omega\T\omega-k_I(F_I-\omega)\T(F_I-\omega)\leq0.
\end{align}

Considering equations (\ref{eq:LypFcn}) and (\ref{eq:VdotLess0}) and evoking the invariance-like theorem 8.4 in~\cite{khalil2002nonlinear} (which uses Barbalat's lemma), we can conclude the following. 
 As $t \rightarrow \infty$, both $\omega$ and $(F_I-\omega)$ approach $0$. Therefore, as $t \rightarrow \infty$, $F_I \rightarrow 0$. In addition, we can conclude that as $t \rightarrow \infty$, $\lan K,I-Q\ran\rightarrow 0$ and hence $Q\rightarrow I$. Therefore the tracking errors 
 $(Q,\omega)$ converge to $(I,0)$ in an asymptotically stable manner.
This means that the proposed control law in equation (\ref{eq:tau}) leads to asymptotically stable tracking of the desired attitude trajectory $(R_d,\Omega^d)$.\qedsymbol\\

\subsection{Robustness to disturbance torque}
The stability result of Theorem 1 guarantees asymptotic convergence of tracking errors $(Q,\omega)$ to $(I,0)$ when there is no disturbance. If there exist a bounded disturbance torque $D$, tracking errors will converge to a bounded neighborhood of $(I,0)$. Theorem 2 gives a specific relation between the size of a neighborhood of $(I,0)$ and disturbance torque $D$, to guarantee convergence of tracking errors $(Q,\omega)$ to that neighborhood of $(I,0).$ 
\\ \\
\noindent \emph{Theorem 2} \\
\emph{
Let $D$ be a disturbance torque that is bounded in norm by a scalar $\gamma$, i.e., $\|D\| \le \gamma$, acting on the dynamics of the system given by equation (\ref{Dyn2}), as follows:
\begin{align}\label{DynDist}
J\dot{\Omega}&=J\Omega \times \Omega+\tau+D.
\end{align} 
Then with the control law given by equation (\ref{eq:tau}), the tracking errors $(Q,\omega)$ converge to a neighborhood of $(I,0)$ given by \begin{align}\label{eq:Boundgamma}
 \mathcal{N}_{(I,0)}:=\{(Q,\omega)\phantom{x}: \phantom{x} \|(2\omega-{F_I})\| \phantom{x} \gamma \leq k_{DI}{\|\omega\|}^2+k_I{\|F_I-\omega\|}^2 \},
\end{align}
asymptotically in time.}
\\  \\
\emph{Proof} \\
By considering disturbed dynamics in equation (\ref{DynDist}) and following similar steps as in the proof of theorem 1, it can be verified that for the disturbed system $\dot{V}$ becomes:
\begin{align}\label{eq:VdotLess0Dist}
   \dot{V}=
   -k_{DI}\omega\T\omega-k_I(F_I-\omega)\T(F_I-\omega)+(2\omega-{F_I})\T D.
\end{align}
The $\phantom{x} (2\omega-{F_I})\T D \phantom{x}$ term is upper bounded by 
\begin{align}\label{eq:VdotLess0Dist3}
   (2\omega-{F_I})\T D\leq \|(2\omega-{F_I})\| \phantom{x} \|D\|\ \leq \|(2\omega-{F_I})\| \phantom{x}\gamma,
\end{align} 
and hence, $\dot{V}$ is upper bounded by
\begin{align}\label{eq:VdotLess0Dist4}
   \dot{V}\leq
   -k_{DI}\omega\T\omega-k_I(F_I-\omega)\T(F_I-\omega)+\|(2\omega-{F_I})\| \phantom{x}\gamma.
\end{align}
Therefore, 
\begin{align}\label{eq:VdotLess0Dist7}
   \dot{V}\leq
   -k_{DI}{\|\omega\|}^2-k_I{\|F_I-\omega\|}^2+\|(2\omega-{F_I})\| \phantom{x}\gamma.
\end{align}
Consequently, $\dot{V}$ is negative semi-definite if  
\begin{align}\label{eq:VdotLess0Dist8}
   -k_{DI}{\|\omega\|}^2-k_I{\|F_I-\omega\|}^2+\|(2\omega-{F_I})\| \phantom{x}\gamma \leq 0,
\end{align}
and that is a sufficient condition for asymptotic convergence of $(Q,\omega)$ to the neighborhood $\mathcal{N}_{(I,0)}$. The size of this neighborhood is given by (\ref{eq:Boundgamma}). \qedsymbol \\

For numerical simulations in the next section, we introduce a time-varying disturbance and show how the proposed PID-type controller effectively compensates for disturbance, compared to a geometric PD-type controller. In addition, we compare our geometric PID-type controller with a classic "non-geometric" PID controller.
\section{Numerical Simulation}\label{sec:NumRes}
For simulation purposes, we consider a quadrotor UAV. The complete control of a quadrotor UAV has two loops: the outer loop position control (translational) and the inner loop attitude control (rotational). The attitude should change such that the desired position trajectory is achieved. In this paper we are looking at the inner loop of attitude control, and proposing a geometric PID type controller for it. However, to have meaningful simulations, we would like to utilize a position controller as well, in conjunction with our proposed attitude controller. Hence for the outer loop of position, we use the translational controller in equation (18) of~\cite{Sasi2018}, as given below:
\begin{align}\label{PositionCon}
    \mathrm{f}= e_3\T R\T\big(mge_3+P\tilde b+L_v (R\nu-v_d)-m\dot v_d\big).
\end{align}
To make meaningful comparisons, we use this outer loop position controller in all the following simulations, while varying the inner loop attitude controller for our comparison purposes. 
In addition, a desired attitude trajectory based on the quadrotor UAV's position trajectory is considered, and then tracked by the proposed algorithm. The desired position trajectory is a helix; going up in the $z$ direction, as shown with a black dotted line in
Fig. \ref{3D tracking}. 

\subsection{Discretization by LGVI}
Discretization of equations of motion is done by utilizing a Lie Group Variational Integrator (LGVI), which, in contrast to general purpose numerical integrators, preserves the structure of configuration space without parameterization or re-projection. The LGVI scheme used in this work was first proposed in~\cite{LGVISE3}. The time step for discretization is a constant $h = t_{k+1}-t_k$. Here $(.)_k$ denotes a parameter of the system at time step $k$. 
The discrete equations of motions are:     
\begin{align*}
R_{k+1}&=R_k\,\mathcal{F}_k,\\
\Omega^{\times}_{d_{k+1}}&=\frac{1}{h}\text{log}(R\T_{d_{k}}R_{d_{k+1}}),\\
J\,\Omega_{k+1}&=\mathcal{F}\T_k\,J\Omega_{k}+h\tau_k,
\end{align*}
where $\mathcal{F}_k \approx \exp(h\Omega^{\times}_k) \in \SO$ is evaluated using Rodrigues' formula:
\begin{align}
   \mathcal{F}_k= \exp({f_k}^{\times})=I +\frac{\sin \|f_k\|}{\|f_k\|}f_k^{\times}+\frac{1-\cos \|f_k\|}{{\|f_k\|}^2}(f_k^{\times})^2,
\end{align}
where
\begin{align}
    f_k=h\Omega_k.\nn
\end{align}
This guarantees that $R_k$ evolves on $\SO$.  For more details on discretization using LGVI please refer to~\cite{LGVISE3}.
\subsection{Simulation Results}\label{DistResults}
\begin{figure}[H]
\centering
\includegraphics[width=.95\columnwidth]{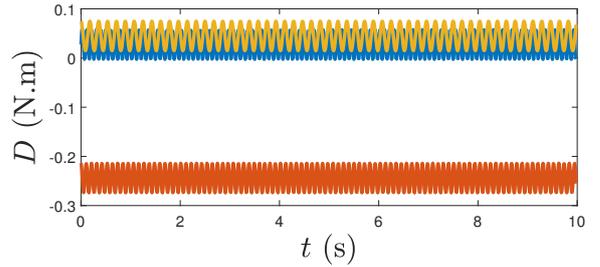}
\caption{Components of disturbance torque $D$ and how they change with time}
\label{Fig:Dist}
\end{figure}

The quadrotor model considered in the simulations has the following physical properties:
\begin{align}
J=diag(0.0820, 0.0845, 0.1377),         
\phantom{x}m=4.34  \text{ kg},
\end{align}
and the time step in simulations is $h=0.01$s. To demonstrate the performance of the proposed geometric PID-type controller, and show effectiveness of the novel integrator term, we introduce a time-varying disturbance torque to the system with components shown in Fig. \ref{Fig:Dist}, and then compare the results with the geometric PD-type attitude controller of~\cite{Sasi2018}. 
The disturbance shown in the above figure consists of the sum of constant and sinusoidal terms. The simulations 
were done using Matlab to encode the LGVI algorithm and the control laws. 
\begin{figure}[H]
\centering
\includegraphics[width=.95\columnwidth]{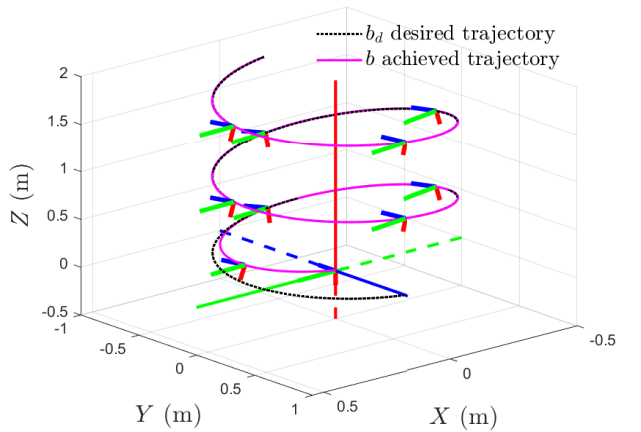}
\caption{3D tracking}
\label{3D tracking}
\end{figure}
Figure \ref{3D tracking} shows how the position and attitude trajectories converge to the desired trajectory using 
the proposed PID-type attitude tracking control in conjunction with the position tracking controller in equation (\ref{PositionCon}).
 
\begin{figure}[H]
\centering
\includegraphics[width=.95\columnwidth]{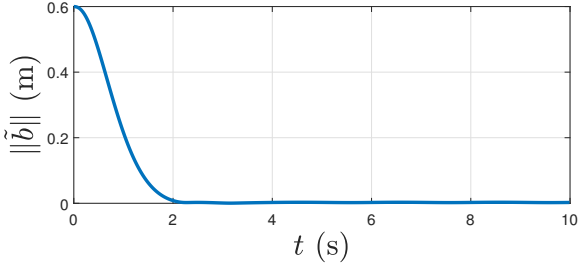}
\caption{Position error norm vs Time}
\label{btildenorm}
\end{figure}
Figure \ref{btildenorm} shows the magnitude of position tracking error over time, and how it converges to zero.

\begin{figure}[H]
\centering
\includegraphics[width=.95\columnwidth]{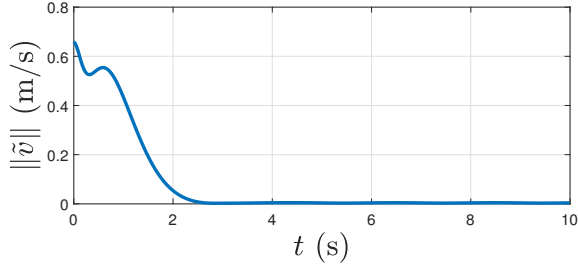}
\caption{Velocity error norm vs Time}
\label{vtildenorm}
\end{figure}
\begin{figure}[H]
\centering
\includegraphics[width=.95\columnwidth]{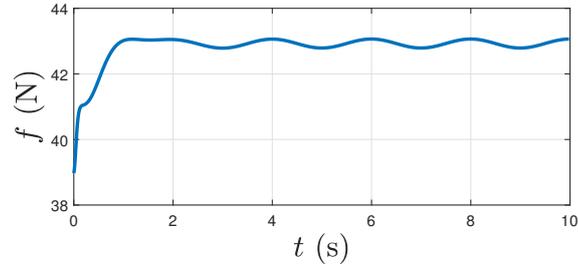}
\caption{Thrust vs Time}
\label{thrust}
\end{figure}
Figure \ref{vtildenorm} shows associated velocity tracking error magnitude converging asymptotically to zero as expected. Figure \ref{thrust} shows the thrust magnitude 
required to track the desired position trajectory.
\begin{figure}[H]
\centering
\includegraphics[width=.95\columnwidth]{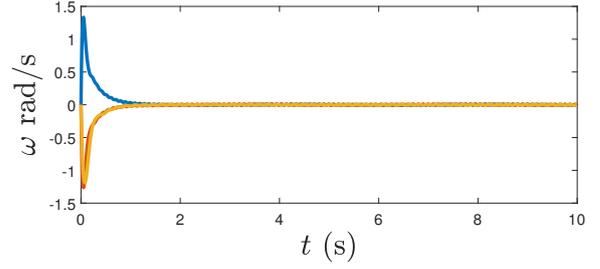}
\caption{Angular velocity tracking error dynamics vs time}
\label{omega}
\end{figure}

Figure \ref{omega} shows components of the angular velocity tracking error, $\omega$, and that they converge 
asymptotically to zero with time. 
\begin{figure}[H]
\centering
\includegraphics[width=.95\columnwidth]{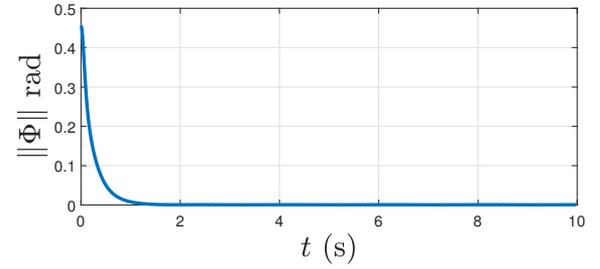}
\caption{Attitude tracking error vs time}
\label{phinorm}
\end{figure}
Figure \ref{phinorm} shows the asymptotic convergence of the attitude tracking 
error with time.

\begin{figure}[H]
\centering
\includegraphics[width=.95\columnwidth]{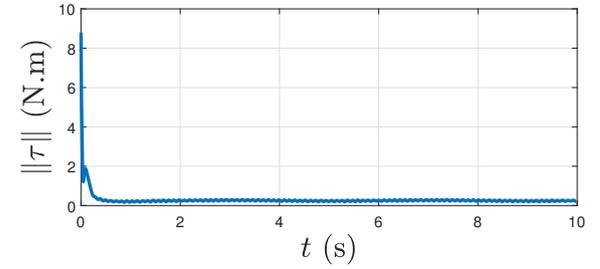}
\caption{Magnitude of Control torque vs time}
\label{torqqueNorm}
\end{figure}
Figure \ref{torqqueNorm} shows the magnitude of the proposed control torque over time.

\subsection{Effectiveness of the integrator term}
Effectiveness of the integrator term in the proposed geometric PID-type attitude control is shown by the following 
comparison:\\
Under the same disturbance torque of Fig. \ref{Fig:Dist}, we use the geometric PD-type attitude controller 
of~\cite{Sasi2018} to track the same trajectory, and compare the results with our proposed geometric PID-type attitude controller.

\begin{figure}[H]
\centering
\includegraphics[width=.95\columnwidth]{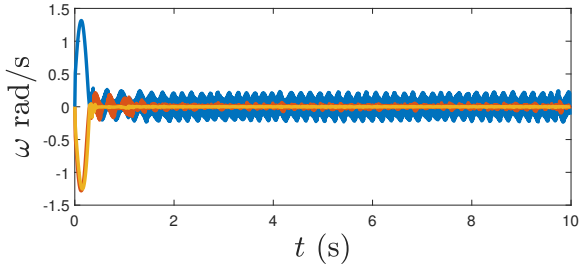}
\caption{Angular velocity tracking error dynamics vs time for PD-type controller of \protect\cite{Sasi2018}}
\label{PDomega}
\end{figure}
Figure \ref{PDomega} shows that for the PD-type controller, components of $\omega$ do not converge to zero but oscillate with noticeable amplitudes about it. On comparing this figure with Fig. \ref{omega}, we see that 
the geometric PID-type controller shows significantly better performance.


\begin{figure}[H]
\centering
\includegraphics[width=.95\columnwidth]{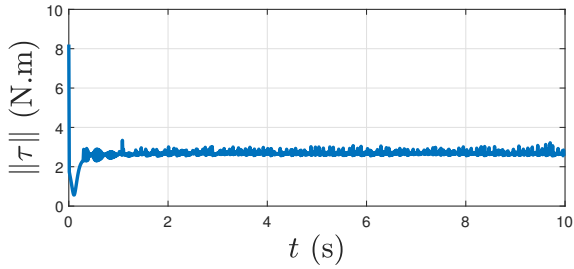}
\caption{Magnitude of Control torque vs time for PD-type controller of \protect\cite{Sasi2018}}
\label{PDtorqqueNorm}
\end{figure}
Figure \ref{PDtorqqueNorm} shows rapid oscillations in control that will likely not be realizable by, and 
therefore should not be implemented on, a quadrotor UAV. Comparing Fig. \ref{torqqueNorm} with Fig. \ref{PDtorqqueNorm}, 
the geometric PID-type controller shows remarkably better performance with negligible oscillations. 
In addition, Figure \ref{PDtorqqueNorm} shows the large magnitude of the required control torque given by the geometric PD-type controller. 
Comparing Fig. \ref{torqqueNorm} with the Fig. \ref{PDtorqqueNorm}, we can see much less required control effort was needed by the 
geometric PID-type controller compared to the geometric PD-type controller.

\begin{figure}[H]
\centering
\includegraphics[width=.95\columnwidth]{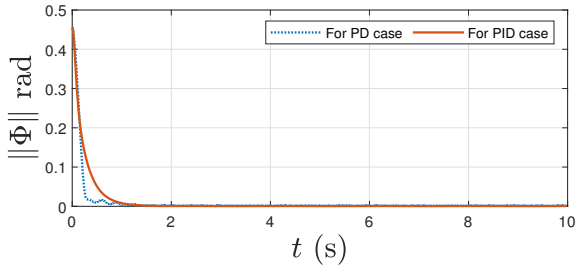}
\caption{Norm of attitude tracking error for the proposed PID-type controller and the PD-type controller in \protect\cite{Sasi2018}}
\label{PDvsPID}
\end{figure}
Figure \ref{PDvsPID} shows norm of the attitude tracking error for both PD and PID-type controllers. The PID-type controller shows this error to be decreasingly more smoothly and with lesser oscillations than the PD-type controller.

Overall, the comparison in this subsection shows that the proposed PID-type controller has significant advantages over a PD-type controller in steady state performance as well as disturbance attenuation. It tracks the same 
maneuvering attitude trajectory better while requiring significantly less overall control effort under the influence of 
a time-varying disturbance torque.

\subsection{Comparing with zero disturbance case}
Another interesting observation can be made by comparing the performance of the proposed PID-type controller under influence of disturbance, which was presented in section \ref{DistResults}, with its performance when there 
is no disturbance ($D=0$), while tracking the same desired trajectory. If disturbance is zero, then the proposed 
PID controller, gives the following results in our numerical simulation.

\begin{figure}[H]
\centering
\includegraphics[width=.95\columnwidth]{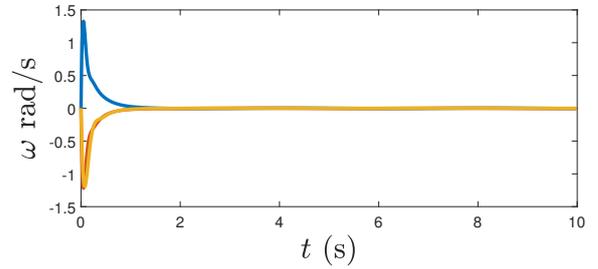}
\caption{Angular velocity tracking error dynamics vs time for $D=0$}
\label{2omega}
\end{figure}

Figure \ref{2omega} shows components of the angular velocity tracking errors over time and that they converge 
asymptotically to zero. This case ($D=0$) can be thought of as an ideal case, and by comparing Fig. \ref{omega} with the above figure, we can see how little the performance of the PID-type controller changes under the influence of disturbance. The time plots in Fig. \ref{omega} show a similar tracking profile to the ideal case of zero disturbance (Fig. \ref{2omega}). 
\begin{figure}[H]
\centering
\includegraphics[width=.95\columnwidth]{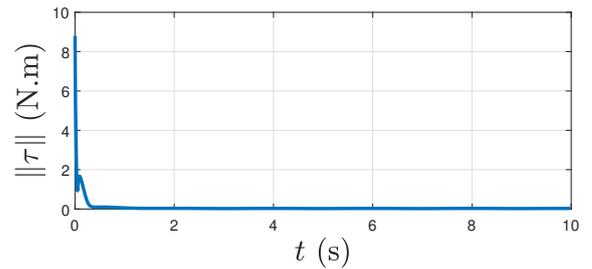}
\caption{Magnitude of Control torque vs time for $D=0$}
\label{2torqqueNorm}
\end{figure}
Figure \ref{2torqqueNorm} shows the control torque magnitude, as given by the proposed PID-type controller when there is no disturbance, i.e., $D=0$. Again  considering this case as an ideal case, we see how well the PID-type controller performs in the presence of disturbance: the profile in Fig. \ref{torqqueNorm} is similar (with minor oscillations) to 
that of Fig. \ref{2torqqueNorm}.\\

To summarize, in this subsection we compared the performance of the PID-type attitude tracking controller for two situations; one when there is no disturbance torque (ideal case) and the other in the presence of a disturbance 
torque. We observed that under a disturbance torque the proposed controller performs well, and results in similar 
(but slightly degraded) performance as in the ideal case of zero disturbance.\\


 \subsection{Comparing with a classic non-geometric PID controller}
 In this part, we utilize a classic non-geometric PID controller given in~\cite{plainPID} as the attitude controller. Following are the results of this simulation. As can be seen from the diverging errors, this non-geometric PID does not perform well with our LGVI dynamics engine.
 \begin{figure}[H]
\centering
\includegraphics[width=.95\columnwidth]{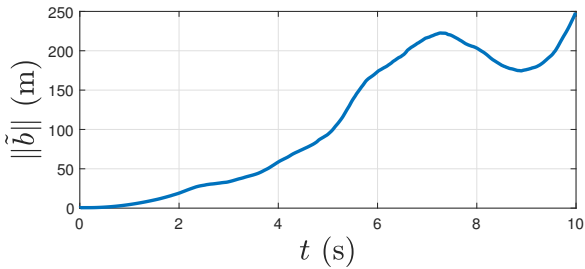}
\caption{Position error norm vs Time for non-geometric PID}
\label{nb10}
\end{figure}
 \begin{figure}[H]
\centering
\includegraphics[width=.95\columnwidth]{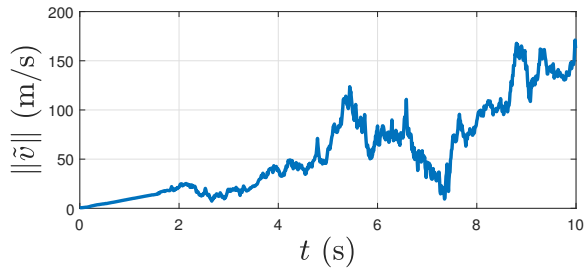}
\caption{Velocity error norm vs Time for non-geometric PID}
\label{nv10}
\end{figure}
 In Figures \ref{nb10} and \ref{nv10} diverging position error norm and velocity error norm are shown, respectively.

 \begin{figure}[H]
\centering
\includegraphics[width=.95\columnwidth]{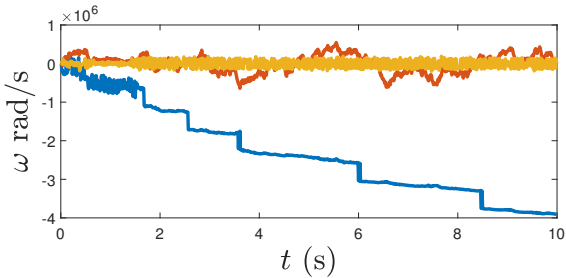}
\caption{Angular velocity tracking error dynamics for non-geometric PID}
\label{Om10}
\end{figure}
 \begin{figure}[H]
\centering
\includegraphics[width=.95\columnwidth]{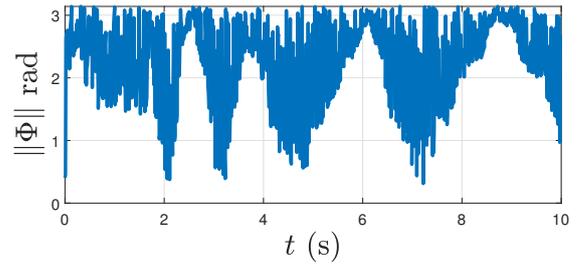}
\caption{Attitude tracking error for non-geometric PID}
\label{nphi10}
\end{figure}
Figures \ref{Om10} and \ref{nphi10} show attitude and angular velocity tracking errors and how they perform.

From this simulation it can be inferred that the geometric PID controller performs better with the LGVI, compared to a classic non-geometric PID controller that can not enforce the errors to zero in conjunction with the position controller in equation (\ref{PositionCon}).
\section{Conclusion and Future Work}\label{sec:conc}
In this work a geometric PID-type attitude tracking control scheme was developed and its stability was shown 
theoretically using Lyapunov analysis on the state space of rigid body attitude motions. Analysis of robustness to disturbance torque was presented. Numerical simulations 
confirmed the performance of the attitude controller, even in the presence of an oscillating disturbance torque, 
in comparison with a geometric PD-type attitude controller, and a non-geometric PID controller. In the near future, we plan to implement this attitude control 
scheme in software-in-the-loop (SITL) simulations using the open source PX4 software for a particular 
quadrotor UAV configuration. This will be followed by hardware implementation on the corresponding 
UAV platform. 
\section*{Acknowledgements}
The authors acknowledge support from the National Science Foundation award CISE 1739748.
\section*{References}
\bibliographystyle{IEEEtran}
\bibliography{eslarticle-template}  

\end{document}